\documentclass[amsmath,amssymb,aps,reprint,superscriptaddress,longbibliography]{revtex4-1}
\usepackage{graphicx}
\usepackage{dcolumn}
\usepackage{bm}

\newcommand\p{\partial}

\newcommand\z{\zeta}

\newcommand\ta{\theta}
\newcommand\ph{\varphi}

\newcommand\eps{\epsilon}

\newcommand\di{\textrm{d}}
\newcommand\ex{\textrm{e}}

\begin{document}

\preprint{APS/123-QED}

\title{Three-dimensional phase field model for actin-based cell membrane dynamics}

\author{Mohammad Abu Hamed}
\affiliation{Department of Mathematics, Technion - Israel Institute of Technology, Haifa 32000, Israel }
\affiliation{Department of Mathematics, The College of Sakhnin - Academic College for Teacher Education, Sakhnin 30810, Israel }

\author{Alexander A. Nepomnyashchy}
\affiliation{Department of Mathematics, Technion - Israel Institute of Technology, Haifa 32000, Israel }

\begin{abstract}
The interface dynamics of a 3D cell immersed in a 3D extracellular matrix is investigated.
We suggest a 3D generalization of a known 2D minimal phase field model suggested in \cite{Ziebert2012}
for the description of keratocyte motility. Our model consists of two
coupled evolution equations for the order parameter and a three-dimensional vector field
describing the actin network polarization (orientation). We derive a closed evolutionary integro-differential equation governing
the interface dynamics of a 3D cell. The equation includes the normal velocity of the membrane, its curvature,
cell volume relaxation, and a parameter  that is
determined by the non-equilibrium effects in the cytoskeleton. This equation can be considered as a 3D generalization
of the 2D case that was derived in \cite{AbuHamed2020}.
\end{abstract}

\maketitle

\section{Introduction}
Over the past decade, increasing attention has been paid to the formulation of computational models that
describe the motility of 3D cells that crawl on flat substrate \cite{Tjhung+Tiribocchi+Marenduzzo+Cates2015},
\cite{Mai+Camley2020}
or invade 3D extracellular matrices (ECM) \cite{Zaman+Kamm+Matsudaira+Lauffenburger2005},
\cite{Schluter+Conde+Chaplain2012}, \cite{Hsun+Gilkes+Wirtz2018}, which is the subject of this paper. This challenge
has been mentioned about a decade ago in the review paper \cite{Mogilner2008} as barely started field.
Since that time, the keratocyte motility, including the lamellipodium waves dynamics
\cite{Mogilner+Barnhart+Keren2020}, has been rather well explored using experimental and theoretical approaches.

Actin polymerization is a basic mechanism in 2D cell motility \cite{Keren2008}, \cite{Mogilner+Barnhart+Keren2020}.
When investigating  the dynamics of 3D cell invasion in 3D ECM, one finds that there are several mechanisms or modes
of
migration that control 3D cell motility, and a cell may switch between them depending on cell intrinsic and extrinsic
factors \cite{Caswell+Zech2018}. Among these mechanisms is the generation of protrusive force through hydrostatic
pressure. This mechanism is called \textit{amoeboid} and it does not need actin polymerization at the leading edge to
generate protrusions.
 Another mechanism, which is the subject of this paper, is the Lamellipodium-based protrusion in 3D-ECMs which
is called
\textit{mesenchymal}. This mechanism is based on actin polarization to form protrusion that enables cell migration in
3D
ECMs, and it is analogous to the 2D actin-based cell motility. This migratory mode has been observed in some
metastatic
cancer cells while moving in complex 3D environments, see \cite{Caswell+Zech2018} and references therein.

Recently a minimal computational phase field model of 3D cell crawling on general substrate topography,  not only on flat surface, has been formulated \cite{Winkler+Aranson+Ziebert2019}.
 In their model they assume that the actin exists only nearby the substrate surface and vanishes far away. Therefore this model is a nontrivial generalization of their 2D model \cite{Ziebert2012}.

In the present paper we suggest a 3D generalization of the original 2D model developed by Aranson and
co-workers \cite{Ziebert2012}, \cite{Ziebert2014} (see also \cite{AbuHamed2020}). It is assumed that the
actin exists over all the interface of the 3D cell, hence that model can and be used as a model of 3D cell surrounded by 3D
ECM.

In the framework of our model, the cell geometry is described by a phase field
coupled with a three-dimensional vector field of the actin network polarization. Thus, we suggest a model of
actin-based 3D cell motility  surrounded by 3D ECM. Relative to other phase field approaches, this model can be
considered as a simple
minimal model describing the 3D cell motility \cite{Ziebert2016}, \cite{Winkler+Aranson+Ziebert2019}.

In the present paper, we consider the case where the cell does not move as a
whole but can change its shape.
The structure of the paper is as follows: In Sec. II we present the
minimal 3D phase field model. In Sec. III we investigate the dynamics of the spherical
shape interface. In Sec. IV we consider the general shape interface. We
derive a closed evolutionary nonlocal equation that describes the interface dynamics. Also, we investigate
the stability of the spherical shape membrane.  Finally, in Sec. V we present the conclusions.

\section{Formulation of the nonlocal problem}
We consider the following minimal phase-field model of self-polarization and motility of spherical cell shape. This
model is a 3D generalization of the 2D model developed in the context of keratocyte motility in
\cite{Ziebert2012} and \cite{Ziebert2014} (see also \cite{AbuHamed2020}).
\begin{subequations}
\begin{eqnarray}
&& u_t = D_u \nabla^2 u  -(1-u)(\delta-u)u -\alpha \nabla u \cdot \textbf{P},\label{Mod1} \\
&& \delta = \frac{1}{2} + \mu  \tilde{V} -\sigma |\textbf{P}|^2,\ \tilde{V}(t)= \int u \di^3 r - V_0 \label{Mod2}\\
&& \textbf{P}_t = D_p \nabla^2 \textbf{P}  - \tau^{-1}\textbf{P}  -\beta \nabla u ,\label{Mod4}\\
&& u(r\rightarrow\infty)=0, \quad 0<u<1,\label{Mod5}\\
&& \textbf{P}(r\rightarrow\infty)=0, \label{Mod6}
\end{eqnarray}
\end{subequations}
where $u(r,\ta,\ph,t)$ is the order parameter that is close to $1$ inside the cell and $0$ outside. The
interface is defined by the relation $u(r=\rho(\ta,\ph,t))=1/2$.
The three-dimensional polarization vector field $\textbf{P}(r,\ta,\ph,t)=p\hat{r}+q\hat{\ta}+w\hat{\ph}$ represents
the actin orientations. It is assumed that $\textbf{P}(r=0)$ is close to zero, and the cell does not move as a whole.
See \cite{AbuHamed2020}
for more details about the formulation of this simplified version of the full model that was developed in \cite{Ziebert2012}.

The model contains several constant parameters: $D_u$ is the stiffness of diffuse interface, $D_p$ is the diffusion
coefficient for $\textbf{P}$, $\alpha$ is the coefficient characterizing advection of $u$ by $\textbf{P}$, $\beta$ determines the creation of $\textbf{P}$ at the interface, $\tau^{-1}$ is the inverse time of the degradation of $\textbf{P}$ inside the cell, $V_0= 4\pi \rho_0^3 /3$ is the overall volume of the cell, $\mu$ is
the stiffness of the volume constraint, and $\sigma$ is the contractility of actin filament bundles. All the parameters listed above are positive.  Notice
that the model (\ref{Mod1})-(\ref{Mod6}) is nonlocal due to the definition of $\delta$.

Because of the spherical symmetry of the problem, we employ the spherical coordinate system with the corresponding differential operators (see Fig. \ref{schematic}),
\begin{subequations}\label{tensors}
\begin{eqnarray}
&&(r,\ta,\ph), \ 0<r<\infty, \ -\frac{\pi}{2}<\ta<\frac{\pi}{2}, \ 0<\ph<2\pi,\\
&& x=r\cos\ta \cos\ph, \ y= r\cos\ta \sin\ph, \ z=r\sin\ta,\\
&& \nabla = \hat{r}\p_r + \hat{\ta}\frac{\p_\ta}{r} + \hat{\ph} \frac{\p_\ph}{r\cos\ta},\\
&& \nabla^2 = \p_{r}^2 + \frac{2\p_r}{r}+ \frac{1}{r^2}\Big( \p_{\ta}^2 -\tan\ta \p_\ta + \frac{\p_\ph^2}{\cos^2 \ta} \Big),\\
&& \nabla^2 \textbf{P} = \hat{r}\nabla^2 p + \hat{\ta}\nabla^2 q + \hat{\ph}\nabla^2 w + O\left(\frac{1}{r^2}\right).
\end{eqnarray}
\end{subequations}
In the next sections we use arguments similar to those we used in the 2D case \cite{AbuHamed2020}.
We begin our analysis with consideration of the spherical cell dynamics.

\section{Rotationally symmetric case}\label{RSC}
Assume that our fields have a rotational symmetry, i.e., $u=u(r,t)$, and $\textbf{P}=p(r,t)\hat{r}$.

In the present paper, the basic assumption is that the ratio $\epsilon$ of the thickness of the cell wall (i.e., the width of the
transition zone, where $u(r,t)$ is changed from nearly 1 to nearly 0) to the size of the cell is small, see Fig. \ref{schematic}.
In that case, the nonlocal term in \eqref{Mod2}  can be estimated as
\begin{equation*}\label{}
   \int u \di^3 r  \approx  \frac{4\pi}{3} \rho^3 (t).
 \end{equation*}
 The model (\ref{Mod1})-(\ref{Mod6}) takes the form,
\begin{subequations}
\begin{eqnarray}
&& u_t = D_u ( u_{rr}+\frac{2}{r}u_r )  -(1-u)(\delta-u)u -\alpha  u_r \cdot p ,\label{rMod1} \\
&& \delta(r,t) = \frac{1}{2}  + \frac{4\pi\mu}{3}( \rho^3 (t) -\rho_0^3  ) -\sigma p^2,\label{rMod2}\\
&& p_t = D_p \left( p_{rr}+\frac{2}{r}p_r -\frac{2}{r^2}p \right)  - \tau^{-1} p  -\beta  u_r ,\label{rMod3}\\
&& u(r\rightarrow\infty)=0, \quad 0<u<1,\label{rMod4}\\
&&  p(r\rightarrow\infty)=0, \quad |p|<1,\label{rMod5}
\end{eqnarray}
\end{subequations}
We introduce the scaling that describes slow dynamics of large enough cell radius,
\begin{equation}\label{scaling}
  \tilde{t}=\epsilon^2 t, \quad \rho(t)=\epsilon^{-1} R(t),
\end{equation}
and define the transition zone variable,
$$\z=r-\rho(t)=O(1).$$
Also we define,
\begin{equation*}\label{}
  R(t) = \tilde{R}(\tilde{t}) , \quad u(r,t) = \tilde{u}(\z,\tilde{t}), \quad p(r,t)=\tilde{p}(\z,\tilde{t}).
\end{equation*}
Consequently the chain rule yields
\begin{equation*}\label{}
  \p_t = -\eps  \tilde{R}_{\tilde{t}}\p_\z + \eps^2 \p_{\tilde{t}}, \quad \p_r = \p_\z.
\end{equation*}
It holds  that
 \begin{equation*}\label{}
  \frac{1}{r} = \frac{\epsilon}{R(t)} - \frac{\epsilon^2 \z}{R^2 (t)} +...
\end{equation*}
The scaled equations \eqref{rMod1}-\eqref{rMod5} are,
\begin{subequations}\label{scal-eq1}
\begin{eqnarray}
&& -\eps R_t u_\z = D_u \left(u_{\z\z} + \frac{2\eps}{R(t)} u_\z \right) -(1-u)(\delta-u)u \nonumber\\
&&  -\alpha u_\z p + O(\eps^2),\label{sc-eq}\\
&& \delta = \frac{1}{2} + \frac{4\pi\mu}{3}\eps^{-3} ( R^3 (t) - R_0^3 )- \sigma p^2 = \frac{1}{2} +  \delta_1,\label{delta-r}\\
&& -\eps R_t p_\z = D_P \left( p_{\z\z} + \frac{2\eps}{R(t)} p_\z \right) -  \tau^{-1} p \nonumber \\
&& -\beta u_\z + O(\eps^2),\\
&& u (\z\rightarrow-\infty)=1, \quad u (\z\rightarrow \infty)=0, \nonumber \\
&& p (\z\rightarrow \pm\infty)=0.
\end{eqnarray}
\end{subequations}
Because the motion of the front is influenced by its curvature $1/R$, the term $R_t u_\z$ in \eqref{sc-eq} should balance the curvature term $2u_\z/R(t)$ in the same equation. That justifies the choice of the time and radius scaling \eqref{scaling}.

\begin{figure}
  \centering
  \includegraphics[scale=0.3]{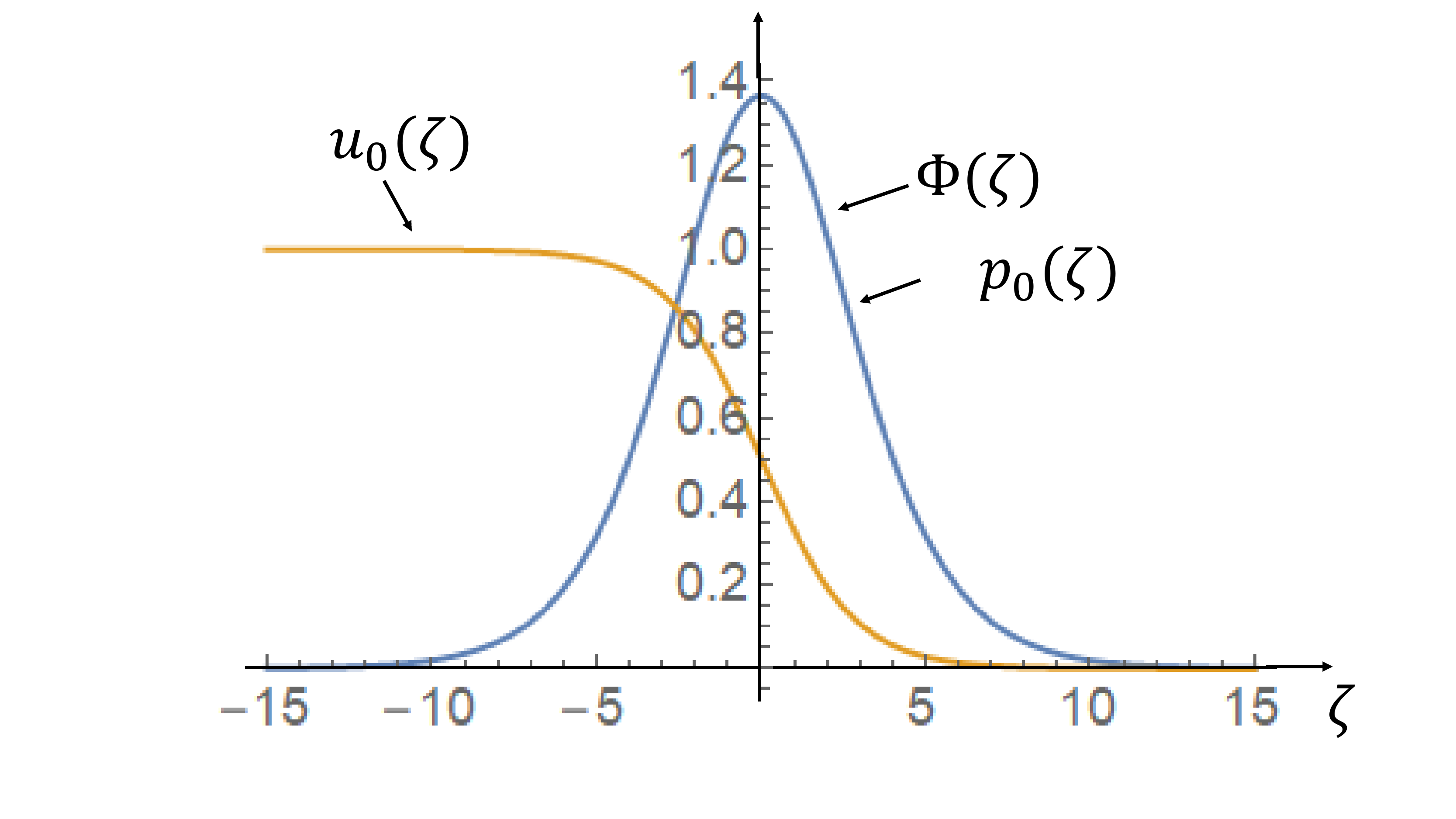}
  \caption{ The plot of the functions $\Phi(\zeta)$ (also of $p_0(\zeta)$ when $\beta=1$) and $u_0(\zeta)$ that describe
the polarization field in (\ref{p0(z)}), (\ref{Phi(z)}), and the kink solution (\ref{u0(z)}) of the order parameter $u_0(\zeta)$.
We employ the values of  parameters $\tau=10, \ D_u =1, \ D_p =0.2, \ A=1, \ M=0.4, \ S=1.5. $   } \label{Phi-zeta}
\end{figure}

Assume that  the parameters have the scaling
\begin{equation}\label{scal1}
 \alpha = \eps A, \   \frac{4\pi\mu}{3}\eps^{-3} = \eps M, \ \sigma = \eps S,
\end{equation}
and introduce the expansions
\begin{equation}\label{expan}
  u = u_0 + \eps  u_1+..., \quad p = p_0 + \eps p_1 +...
\end{equation}
 Substituting \eqref{scal1} and \eqref{expan} into \eqref{scal-eq1} and collecting terms of the same
order,  we obtain at the leading order system \eqref{led},
\begin{subequations}\label{led}
\begin{eqnarray}
&& D_u u_{0\z\z} = (1-u_0)\left(\frac{1}{2}-u_0 \right)u_0\label{led1}\\
&& D_p p_{0\z\z} -\tau^{-1}p_0 = \beta u_{0\z}\label{led2}\\
&& u_0(\z\rightarrow-\infty)=1, \quad u_0(\z\rightarrow \infty)=0, \\
&& p_0(\z\rightarrow \pm\infty)=0.
\end{eqnarray}
\end{subequations}
Following the Ginzburg-Landau theory and Fourier transform method, we find the solutions of the system,
\begin{subequations}\label{rfield}
\begin{eqnarray}
  &&  u_0(\z) = \frac{1}{2} \left[ 1-\tanh\left(\frac{\z}{\sqrt{8D_u}}\right) \right],\label{u0(z)} \\
  && p_0 (\z) =\beta \Phi(\tau,D_u , D_p, \z  ), \label{p0(z)} \\
  && \Phi(\tau,D_u , D_p, \z  ) =\label{Phi(z)}\\
  &&\frac{1}{8}\sqrt{\frac{\tau}{2 D_u D_p}} \int_{-\infty}^{\infty} \ex^{-|r|/\sqrt{\tau D_p}} \cosh^{-2} \left( \frac{r- \z}{\sqrt{8D_u}} \right) \di r. \nonumber
\end{eqnarray}
\end{subequations}
See Fig. \ref{Phi-zeta} for the plot of the fields in \eqref{rfield}.
The equation for $u_1$ at the order $O(\eps)$ is
\begin{eqnarray*}
&& D_u u_{1\z\z} - \left(\frac{1}{2} -3 u_0 + 3u_0^2 \right)u_1 =\nonumber \\
&& (1-u_0)u_0 \eps^{-1} \delta_1 + A u_{0\z}p_0 - \frac{2D_u}{R(t)}u_{0\z} - R_t u_{0\z}.
\end{eqnarray*}
The solvability condition of the latter equation yields the following closed equation for the interface dynamics
$R(t)$,
\begin{eqnarray}
&& \int_{-\infty}^{\infty} \di \z u_{0\z} \Big\{ - R_t u_{0\z} - \frac{2D_u}{R(t)}u_{0\z}+ A u_{0\z}p_0 + \nonumber \\
&& u_0 (1-u_0) \left[ M \left( R^3 (t) -  R_0^3 \right) - S p_0^2  \right]  \Big\} =0; \label{R(t)0}
\end{eqnarray}
see \cite{AbuHamed2020} for more details.
The expression (\ref{u0(z)}) yields,
\begin{equation}\label{}
\int_{-\infty}^{\infty} \di \z u_{0\z}^2=\frac{1}{6\sqrt{2D_u}}, \quad  \int_{-\infty}^{\infty} \di \z u_{0\z}u_0(1-u_0)=-\frac{1}{6}.
\end{equation}
Hence equation (\ref{R(t)0}) may be written in the form ,
\begin{eqnarray}
&& \frac{1}{\sqrt{2D_u}}\left( R_t + \frac{2D_u}{R} \right) =  M[ R_0^3 - R^3 (t) ]  + \Omega(\beta)\label{R(t)} \\
&&\Omega(\beta) =  6\beta \left(A \Omega_1 -  \beta S \Omega_2  \right) , \label{Omega}
\end{eqnarray}
where,
\begin{subequations}\label{Omega1}
\begin{eqnarray}
&& \Omega_1 (\tau ,D_u , D_p  ) = \int_{-\infty}^{\infty} \Phi(\z) u_{0\z}^2 \di \z, \\
&& \Omega_2 (\tau, D_u , D_p  ) = \nonumber \\
&&\int_{-\infty}^{\infty} \Phi^2 (\z) (1-u_0) u_0 u_{0\z} \di \z<0.
\end{eqnarray}
\end{subequations}
It is more convenient to consider the following presentation of equation (\ref{R(t)}) of the front dynamics,
\begin{subequations}
\begin{eqnarray}
&& a R_t  = \frac{f(R)}{R},\label{R(t)1} \\
&& f(R) = -MR^4 + [ MR_0^3 +\Omega(\beta) ]R -\sqrt{2D_u}, \label{R(t)1b}
\end{eqnarray}
\end{subequations}
where $a=1/\sqrt{2D_u}$. The expression in the right-hand side of (\ref{R(t)1}) has the meaning of a "force" acting on the cell surface. Therefore, equation
(\ref{R(t)1}) can be considered as a ``force-velocity relation" of the motionless 3D cell immersed in 3D ECM (see \cite{Keren2008}).

Recall that $\Omega_1$ is positive and $\Omega_2$ is negative, therefore, $\Omega(\beta)$ is a monotonically growing function of $\beta$, see \eqref{Omega},\eqref{Omega1}. The nonlinearity of the dependence of $\Omega$ on $\beta$ is caused by the nonlinear term $-\sigma p^2$ in the expression \eqref{delta-r} for $\delta$. The physical origin of that term is the contraction of actin
filament bundles.

\begin{figure}
  \centering
  \includegraphics[scale=0.3]{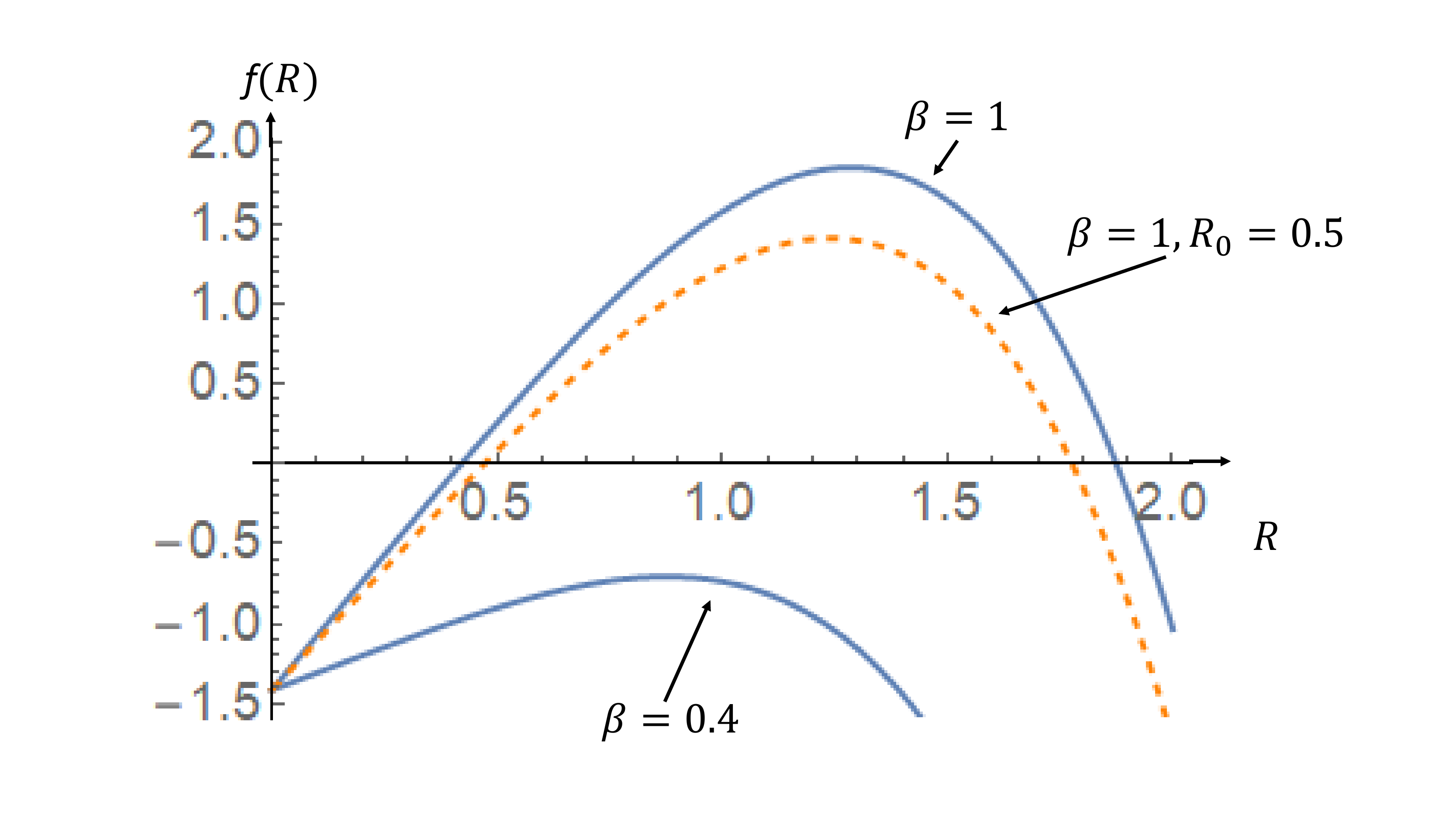}
  \caption{ Solid lines present the plots of the function $f(R)$ in \eqref{R(t)1},\eqref{R(t)1b} for the
values $R_0=1$, and $\beta=1$ that manifest stable and unstable states, and for the value $\beta=0.4$ that is nonphysical since it does not include any steady states.
 We use the same values of parameters as in Fig. \ref{Phi-zeta}. The dashed line is for the
case $\beta=1$, and $R_0=0.5$.   } \label{f-R}
\end{figure}

In  Fig. \ref{f-R} we plot the function $f(R)$ for two values of $\beta$;  the graph shows the existence of
two stationary radii, stable, $R_+(\beta)$ and unstable, $R_-(\beta)$, for $\beta=1$,
while no stationary states for $\beta=0.4$. One can conclude that there exists a critical value $\beta_c$ such that
there are no stationary
solutions when $\beta<\beta_c$. Below we find that $\beta_c=0.637$ for the chosen set of parameters.

\begin{figure}
    \includegraphics[scale=0.3]{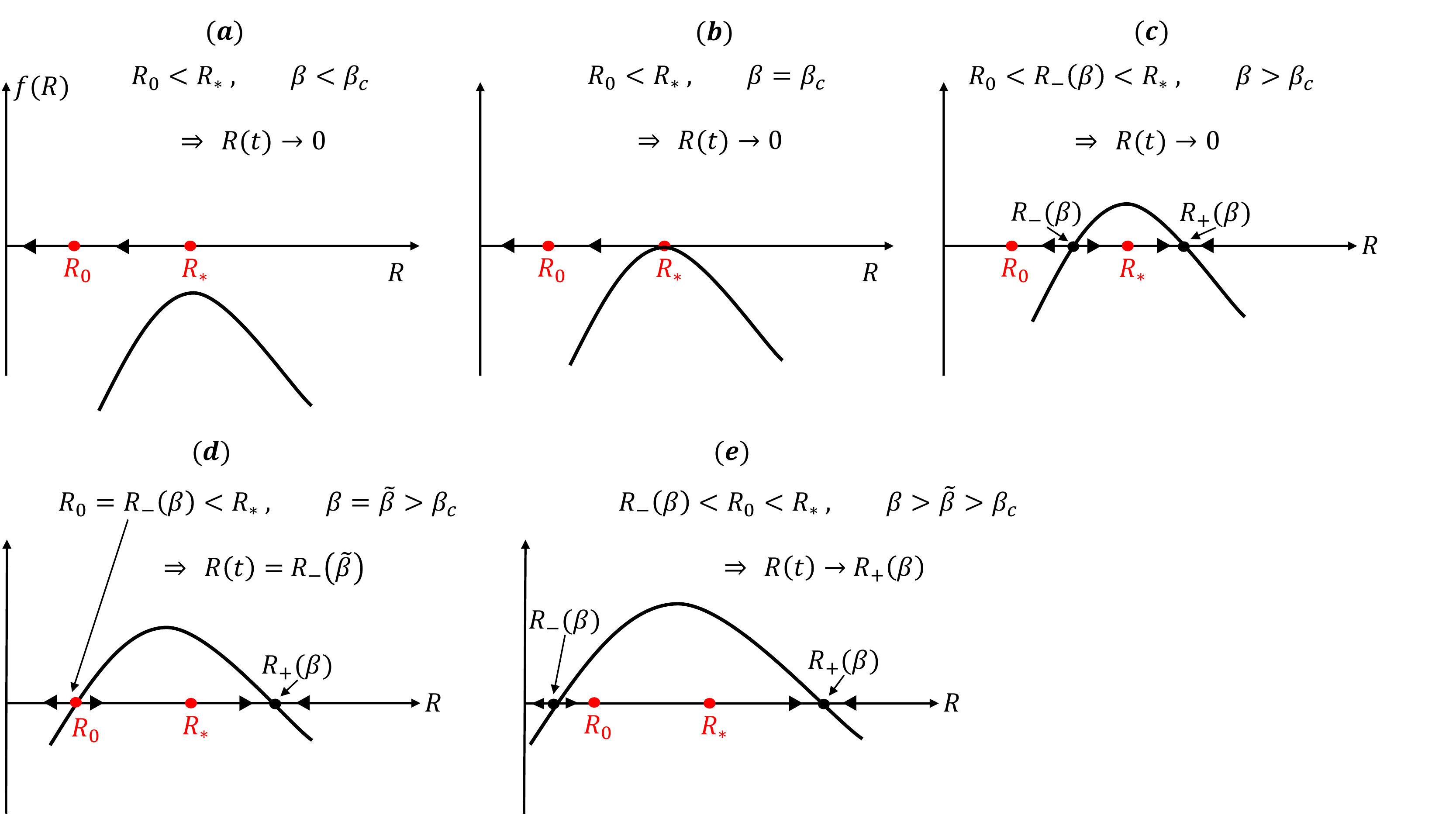}
  \caption{A schematic description of the radius dynamics $R(t)$ that is governed by the ODE (\ref{R(t)1}), for all the subcases of the case $R_0<R_*$ as explained in the text. Notice that the positions of $R_0$, and $R_*$ are fixed (red points) for all the values of $\beta>0$.  } \label{R0-beta}
\end{figure}

The critical value $\beta_c$ has to satisfy three constraints: (i) $MR_0^3 +\Omega(\beta_c) >0$, which
guarantees the existence of maximum of $f(R)$ at a certain $R=R_{*}$, (ii) $f(R_*,\beta_c)=0$, (iii) $f'_R(R_*,\beta_c)=0$ (see Fig. \ref{R0-beta}(b)). As a
result we find that $\beta_c$ is the positive solution of the quadratic equation
  \begin{equation}\label{b-c}
  MR_0^3 +\Omega(\beta_c) = 2^\frac{19}{8} 3^\frac{-3}{4} D^\frac{3}{8} M^\frac{1}{4}   ,
\end{equation}
that can be found explicitly:
\begin{eqnarray*}\label{}
 && \beta_c =  \frac{1}{2S\Omega_2 }\Bigg[ A\Omega_1 -\\
 &&\sqrt{(A\Omega_1)^2-S\Omega_2 \left(2^\frac{27}{8} 3^\frac{-7}{4} D^\frac{3}{8} M^\frac{1}{4} -\frac{2}{3}MR_0^3\right) }\Bigg]
 \end{eqnarray*}
For $\beta=\beta_c$,
$$R_-(\beta_c)=R_+(\beta_c)=R_*=2^\frac{1}{8} 3^\frac{-1}{4} D^\frac{1}{8} M^\frac{-1}{4};$$
notice that $R_*$ does not depend on $R_0$;  for values of parameters indicated in Fig. \ref{Phi-zeta}, $R_*=1.041$ .  Because $\Omega(\beta)$ is a
monotonically growing function of $\beta$, $R_-(\beta)$ decreases with the growth of $\beta$, and therefore $R_-(\beta)<R_*$ for any $\beta>\beta_c$.
On the contrary, $R_+(\beta)$ increases with the growth of $\beta$. For values of parameters indicated in Fig. \ref{Phi-zeta} and $R_0=1$, we find $\beta_c=0.637$ .

If $\beta<\beta_c$, then $f(R)<0$ for any $R$, therefore the cell radius decreases with time and tends to zero during a finite time (see Fig. \ref{R0-beta}(a)). The
temporal evolution of $R(t)$ in the case of $\beta>\beta_c$ depends on the relation between the initial radius $R_0$ and $R_-(\beta)$. If
$R_0>R_-(\beta)$, then $R(t)\to R_+(\beta)$ as $t\to\infty$; if $R_0<R_-(\beta)$, then $R(t)\to 0$ during a finite time. Therefore, if
$R_0>R_*$, $R(t)$ tends to a finite value for any $\beta>\beta_c$, because $R_-(\beta)<R_*<R_0$. However, if $R_0<R_*$, the cell shrinks even at
$\beta>\beta_c$, if $\beta$ is still less than a certain value $\tilde{\beta}$ (see Fig \ref{R0-beta}(c)), which is determined by the relation
$R_0=R_-(\tilde{\beta})$ (see Fig. \ref{R0-beta}(d)).  Because $f(R_0,\tilde{\beta})=0$, the value of $\tilde{\beta}$ can be found by solving the equation,
\[
\Omega(\tilde{\beta})= \frac{\sqrt{2 D_u} }{R_0}.
\]
 For $\beta>\tilde{\beta}$, $R_-(\beta)<R_0$, therefore the cell radius tends to a finite value (see Fig. \ref{R0-beta}(e)).

\begin{figure}
  \centering
  \includegraphics[scale=0.3]{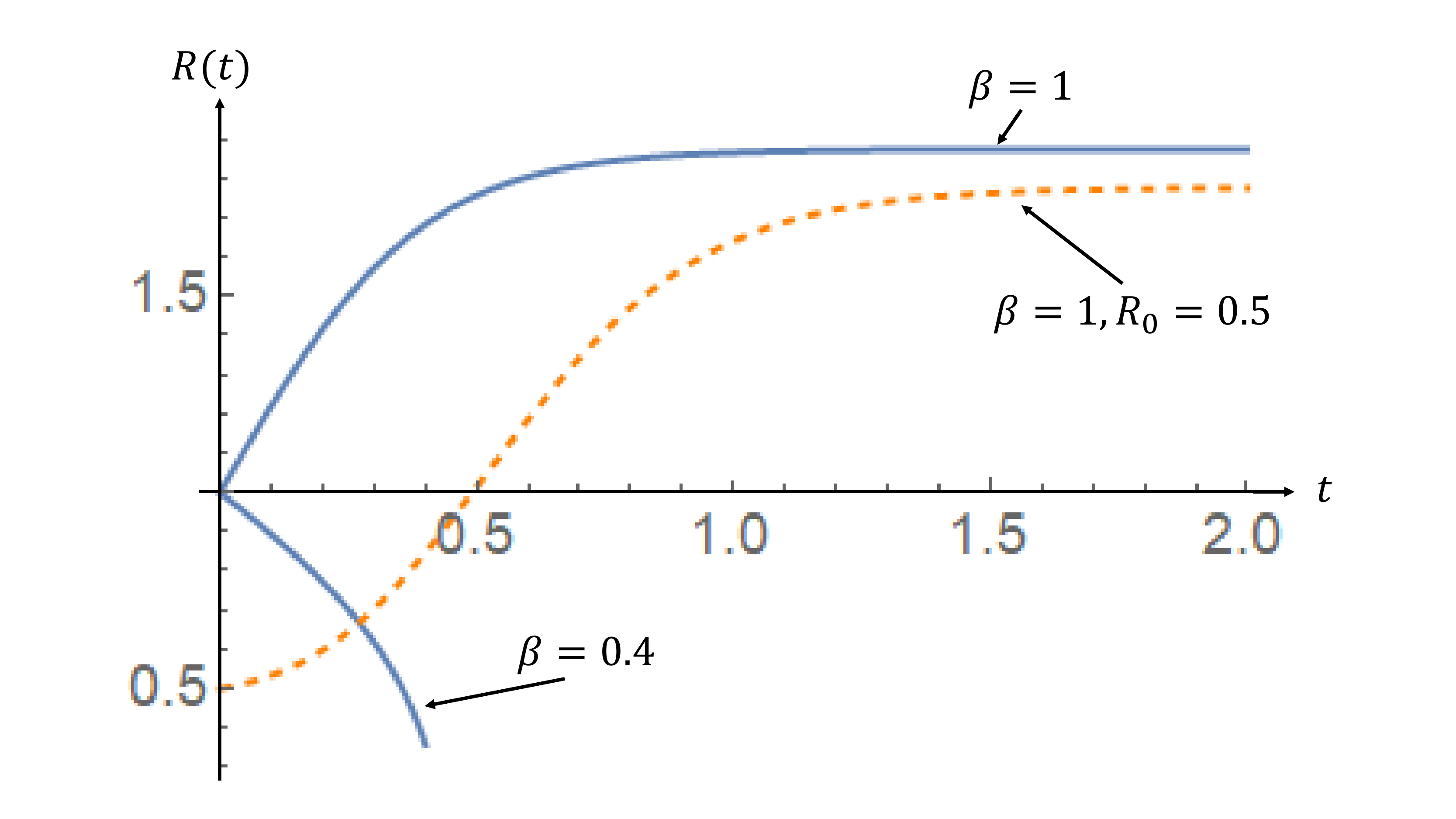}
  \caption{ Solid lines: numerical solutions of the ODE (\ref{R(t)1}) that yields $R(t)$
for values $R_0=1$, $\beta=1$ that manifest the existence of stationary radius,
and for the value $\beta=0.4$ is below the critical value $\beta_c=0.637$ that is meaningless in the context of cell dynamics.  We use the same values of
parameters as in Fig. \ref{Phi-zeta}. The dashed line is for the case $\beta=1$, and $R_0=0.5$. } \label{R-t}
\end{figure}

In Fig. \ref{R-t} we present the numerical solution of the ODE \eqref{R(t)1}
for the  values $\beta=1$ . As we can see, the cell radius can increase  monotonically until it
reaches the steady state value. This is because in the framework of our model, the volume is not conserved and
influences the dynamics through the parameter $\delta(r,t)$, see equation \eqref{delta-r}.

 The decrease of the cell radius (in the language of the phase-field model, the transition of the phase $u=1$ into the phase $u=0$) is caused
by negative terms in the right-hand side of \eqref{Mod1},\eqref{rMod1}, among them the term $-\delta(1-u)u$, which is  negative at large $R$, and
by the diffusion term, which creates an effective ``surface tension" of the cell surface. The positive term $-\alpha u_r p$ hinders the decrease
of the cell radius. If $\beta$ is not sufficiently large, the polarization is not strong enough to stop the collapse of the cell. In that case, the
cell shrinks until it disappears (see Fig. \ref{f-R}, $\beta=0.4$.)

Note that at large $\beta$ (see Fig. \ref{R0-beta}(e)) we have $\Omega(\beta)\gg1$, and $R_-(\beta)\ll1$, therefore one can approximate the expression of $f(R)$ in (\ref{R(t)}) as:
$$R_-(\beta)\approx\frac{\sqrt{2D_u}}{\Omega(\beta)} .$$
We can see that for sufficiently large $\beta$, $R_-(\beta)$
can be arbitrary small. Therefore, for arbitrary small $R_0$, there exists
such $\tilde{\beta}$ that the cell radius tends to a finite value, if $\beta>\tilde{\beta}$.

For values of parameters indicated in Fig. \ref{Phi-zeta} and $R_0=0.5$, $\tilde{\beta}=0.96$ . In Fig. \ref{f-R} and \ref{R-t}, we show the numerical plots for the parameters $\beta=1$, and $R_0=0.5$.

Let us emphasize that the described effect is not physical:
for such values of parameters, the model does not reflect the true
behavior of cells.

\section{Dynamics of general shape interface}
We  employ the scaling and definitions of the previous section. Now we have
to consider the azimuthal dependence of variable. One can
calculate,
\begin{eqnarray*}
&& \p_\ta = -\eps^{-1} R_\ta \p_\z + \p_\ta, \ \p_\ph = -\eps^{-1} R_\ph \p_\z + \p_\ph,\\
&& \p_\ta^2  = \eps^{-2} R_\ta^2 \p_{\z}^2 - \eps^{-1}( R_{\ta\ta}\p_\z +2R_\ta \p_{\z\ta}^2 ) + \p_{\ta}^2,\\
&& \p_\ph^2  = \eps^{-2} R_\ph^2 \p_{\z}^2 - \eps^{-1}( R_{\ph\ph}\p_\z +2R_\ph \p_{\z\ph}^2 ) + \p_{\ph}^2,\\
&& \frac{1}{r}\p_\ta = -\frac{R_\ta}{R} \p_\z +O(\eps),\quad \frac{1}{r^2}\p_\ta = -\eps \frac{R_\ta}{R^2} \p_\z +O(\eps^2),\\
&& \frac{1}{r}\p_\ph = -\frac{R_\ph}{R} \p_\z +O(\eps),\quad \frac{1}{r^2}\p_\ph = -\eps \frac{R_\ph}{R^2} \p_\z +O(\eps^2),\\
&& \nabla^2 u = \left( 1 + \frac{R_\ta^2}{R^2} +\frac{R_\ph^2}{R^2 \cos^2 \ta} \right) u_{\z\z} + \nonumber\\
&& \eps \Bigg[   \left( \frac{2}{R} + \frac{R_\ta}{R^2}\tan\ta - \frac{R_{\ta\ta}}{R^2} - \frac{R_{\ph\ph}}{R^2 \cos^2 \ta }\right)u_\z \nonumber\\
&&  -\frac{2}{R^2}\left( R_\ta u_{\z\ta} + \frac{R_\ph}{\cos^2 \ta}u_{\z\ph} \right) \nonumber \\
&& -\frac{2\z}{R^3} \left(R_\ta^2 + \frac{R_\ph^2}{\cos^2 \ta} \right) u_{\z\z}  \Bigg]+O(\eps^2). \label{LO}
\end{eqnarray*}
The nonlocality in (\ref{Mod2}) is approximated as follows,
 \begin{equation*}\label{}
   \int u \di^3 r \sim   \frac{\eps^{-3}}{3} \int_{0}^{2\pi} \di \ph \int_{-\pi/2}^{\pi/2} R^3 (\ta,\ph,t) \cos\ta  \di \ta .
 \end{equation*}
 We use expansions (\ref{expan}),and define the auxiliary
function $\Lambda$, 
\begin{equation*}\label{}
   \Lambda(\ta,\ph, t)= \left( 1+ \frac{R_\ta^2}{R^2} + \frac{R_\ph^2}{R^2 \cos^2 \ta} \right)^{-1/2} .
 \end{equation*}
 At the leading order one can fined,
 \begin{eqnarray*}\label{}
&& D_u \Lambda^{-2} u_{0\z\z} = (1-u_0)(\frac{1}{2}-u_0)u_0,\\
&& D_p \Lambda^{-2} p_{0\z\z} - \tau^{-1} p_0 = \beta u_{0\z},\\
&& D_p \Lambda^{-2} q_{0\z\z} - \tau^{-1} q_0 = -\beta \frac{R_\ta}{R} u_{0\z},\\
&& D_p \Lambda^{-2} w_{0\z\z} - \tau^{-1} w_0 = -\beta \frac{R_\ph}{R\cos\ta} u_{0\z},
\end{eqnarray*}
therefore similarly to the previous section one can calculate the solutions
\begin{subequations}
 \begin{eqnarray}\label{}
&& u_0 (\z) = \frac{1}{2} \left[ 1-\tanh\left(\frac{ \Lambda\z}{\sqrt{8D_u}}\right) \right], \label{l1}\\
&& p_0 (\z)  = \beta\Lambda  \Phi(\Lambda\z)\label{l2}\\
&& q_0(\z) = -\beta\Lambda  \frac{R_\theta}{R}  \Phi(\Lambda\z)\label{l3}\\
&& w_0(\z) = -\beta\Lambda  \frac{R_\ph}{R\cos\ta}  \Phi(\Lambda\z)\label{l4}
\end{eqnarray}
\end{subequations}
 The equation for $u$ at the order $O(\eps)$ have the form,
\begin{eqnarray}\label{}
&& D_u \Lambda^{-2} u_{1\z\z} - \left(\frac{1}{2} -3 u_0 + 3u_0^2 \right)u_1 = - R_t u_{0\z}+\nonumber\\
&& -D_u \Bigg[ \left( \frac{2}{R} + \frac{R_\ta}{R^2}\tan\ta - \frac{R_{\ta\ta}}{R^2} - \frac{R_{\ph\ph}}{R^2 \cos^2 \ta }\right)u_{0\z} \nonumber\\
&&  -\frac{2}{R^2}\left( R_\ta u_{0\z\ta} + \frac{R_\ph}{\cos^2 \ta}u_{0\z\ph} \right) \nonumber \\
&& -\frac{2\z}{R^3} \left(R_\ta^2 + \frac{R_\ph^2}{\cos^2 \ta} \right) u_{0\z\z}  \Bigg]\label{curef}\\
&& + A  u_{0\z} \left( p_0 -\frac{R_\ta}{R} q_0 - \frac{R_\ph}{R\cos\ta}w_0  \right) +\nonumber\\
&& (1-u_0)u_0 \left\{  \tilde{V}(t) -  S(p_0^2 + q_0^2 + w_0^2)  \right\}, \nonumber
\end{eqnarray}
 where the volume variation has the form,
 \begin{equation*}
  \tilde{V} (t)= M\left[ \frac{1}{4\pi}  \int_{0}^{2\pi} \di \ph \int_{-\pi/2}^{\pi/2} R^3 (\ta,\ph,t) \cos\ta  \di \ta - R_0^3 \right].
 \end{equation*}
 We apply the solvability condition, which is the orthogonality of the
equation's right-hand side to the solution of the homogenous equation
$u_{0\z}$, and obtain a closed form of the interface dynamics,
  \begin{equation}\label{gfd}
   a \Lambda R_t = - 2a D_u \mathcal{H} - \tilde{V} + \Omega,
 \end{equation}
 where
 \begin{equation}\label{curv}
 \mathcal{H} =\frac{1}{2} \nabla\cdot \hat{n} = \frac{1}{2} \nabla\cdot \left( \frac{\nabla(r-R)}{|\nabla(r-R)|} \right)
 \end{equation}
 is the mean local curvature of the surface $r= R(\ta,\ph,t)$. For an explicit expression of the curvature see Appendix \ref{A}.
 Note that in the spherically symmetric case $\Lambda=1$, $H=1/R$ and $\tilde{V}=M[R^3(t)-R_0^3]$.
Therefore, equation \eqref{R(t)} is recovered from equation \eqref{gfd}.
 Notice that $\Omega$ in equation \eqref{gfd}, depends on all of the parameters that describe the nonequilibrium molecular effects of the subcell level, see \eqref{Omega}, and \eqref{Omega1}.
Equation \eqref{gfd} is a closed evolutionary equation for the 3D cell interface dynamics , which is an integro-differential equation, i.e., it is nonlocal, unlike that obtained
in the spherical case \eqref{R(t)1}.
 For the details of the application of the solvability
condition in order to obtain equation \eqref{gfd},  we refer the reader to
\cite{AbuHamed2020} and \cite{AbuHamed2016}, where we perform similar calculations.

Notice that in \eqref{gfd} the expression $\Lambda R_t$ corresponds to the normal velocity $v_n$ of the interface, thus that equation is a generalization of the well-known curvature flow. By a proper scaling transformation, $t\rightarrow a^2 t $ and $R(t)\rightarrow aR(t) $, the equation of motion of the cell boundary can
be brought to a canonical form,
 \begin{equation*}\label{}
   v_n = - 2D_u \mathcal{H} - \tilde{V} + \Omega.
 \end{equation*}
In addition it suggests an answer for the unrevealed force – velocity
relation for the actin network that was highlighted in \cite{Keren2008} in the context of shape dynamics of a 2D cell.

\subsection{Stability of the radial interface}
Let our base radial solution in (\ref{R(t)1}) be perturbed
\begin{equation*}\label{}
   R = \bar{R}(t) + \hat{R}(\ta,\ph,t), \quad \hat{R} \ll \bar{R}.
 \end{equation*}
The linearization of (\ref{gfd}) around that base solution yields,
 \begin{eqnarray}\label{}
&& \hat{R}_t =\frac{ D_u }{\bar{R}^2(t)} \left( 2\hat{R} + \hat{R}_{\ta\ta} - \hat{R}_\ta \tan\ta + \frac{\hat{R}_{\ph\ph}}{\cos^2 \theta}  \right) \nonumber \\
&& - \frac{3M \bar{R}^2}{4\pi a} \int_{0}^{2\pi} \di \ph \int_{-\pi/2}^{\pi/2} \hat{R} \cos\ta  \di \ta. \label{Leq}
\end{eqnarray}
First, let us consider solutions satisfying the condition
\begin{equation}\label{int1}
\int_{-\pi/2}^{\pi/2} \hat{R} \cos\ta  \di \ta=0.
\end{equation}
Let us consider the normal mode $\hat{R}(\theta,\phi,t)=T(t)\Theta(\theta)e^{im\ph}$. For $m\ne0$ the integral term in (\ref{Leq}) vanishes thus we do not have a contribution of the volume variation. For $m=0$ it vanishes due to condition \eqref{int1} . Applying the separation of variable method one obtain,
\begin{eqnarray*}\label{}
  && \frac{\bar{R}(t)}{D}\frac{T'}{T}=\frac{\Theta''-\tan\theta\Theta' + \left(2-\frac{m^2}{\cos^2\theta}\right)\Theta }{\Theta}=-\mu\\
  && \Rightarrow \Theta''-\tan\theta\Theta' + \left(2+\mu-\frac{m^2}{\cos^2\theta}\right)\Theta=0.
 \end{eqnarray*}
 The solution $\Theta$ is bounded if $2+\mu=n(n+1)$; in the latter case we obtain the spherical harmonics solution
with the associated Legendre function,
 \begin{eqnarray*}\label{}
  && \hat{R} = \ex^{im\ph} \mathcal{P}_n^m (-\sin\ta) T_n(t), \\
  && n = 0,1,2,..., \quad m=0,1,..., n. \nonumber
 \end{eqnarray*}
All solutions with $n\neq 0$ satisfy condition \eqref{int1} due to the orthogonality property of the spherical harmonics,
therefore  $T_{n}(t)$ satisfy,
\begin{equation*}\label{}
   T_n'(t)= \frac{[2-n(n+1)]D_u }{\bar{R}^2 (t)} T_n.
 \end{equation*}
  For $n=1$, which corresponds to the spatial translation of the
sphere as a whole, we find that $T_1(t)=const$. Disturbances with $n \geq2$, which describe the shape distortions,
decay with time.

In the case $n=m=0$, which corresponds to a change of the sphere
radius, the integral
$$I=  \int_{-\pi/2}^{\pi/2}\Theta(\theta)\cos\theta \di\theta\neq 0,$$
thus from equation \eqref{Leq}  we obtain the equation,
\begin{eqnarray*}\label{}
  && T'\Theta = \frac{D_u}{\bar{R}^2(t)}T[\Theta''-\Theta'\tan\theta +2\Theta ] -  \\
  && \frac{3M\bar{R}^2(t)}{2a} T \int_{-\pi/2}^{\pi/2} \Theta(\theta)\cos\theta \di \theta.
 \end{eqnarray*}
Let us divide both sides of the equation by $T\Theta$. We can see that both sides of the obtained equality are functions only
of $t$:
\begin{equation*}
  \frac{T^{\prime}}{T}=\frac{1}{\Theta}
  \Big[\frac{D_u}{\bar{R}^2(t)} \left(\Theta^{\prime\prime}-\Theta^{\prime}\tan\Theta+2\Theta\right)- \frac{3M\bar{R}^2(t)}{2a} I\Big]\equiv\lambda(t),
\end{equation*}
hence
\begin{equation}\label{int2}
  \frac{D_u}{\bar{R}^2(t)} \left[\frac{1}{\cos\theta}\frac{d}{d\theta} \left(\cos\theta\frac{d\Theta}{d\theta}\right)+2\Theta\right]-
\frac{3M\bar{R}^2(t)}{2a}I=\lambda(t)\Theta.
\end{equation}
Multiplying both sides of \eqref{int2} by $\cos\theta$, integrating over $\theta$ from $-\pi/2$ to $\pi/2$, and dividing by
$I\neq 0$, we find that
$$\frac{T^{\prime}(t)}{T(t)}=\lambda(t)=\frac{2D_u}{\bar{R}^2(t)}- \frac{3M\bar{R}^2(t)}{a},$$
therefore
\begin{eqnarray}
  && T_0'(t)= \sqrt{2D_u}\frac{\di}{\di R} \left[\frac{f(R)}{R}\right]\bigg|_{R=\bar{R}(t)} T_0= \nonumber\\
  &&  \left[  \frac{2D_u}{\bar{R}^2(t)}  - \frac{3M\bar{R}^2 (t)}{a}\right] T_0.\label{T0}
 \end{eqnarray}
 The analysis of the sign of the expression in the right-hand side of
\eqref{T0} confirms the result obtained in Section \ref{RSC}: solution $R=R_-$ is
unstable and solution $R=R_+$ is stable with respect to the radius
change.  $T_0$ can grow, but when $\bar{R}(t)$ approaches its stationary values, the
derivative $d(f(R)/R)/dr$ at $R=\bar{R}(t)$ becomes negative (see Fig. \ref{f-R}),
hence the spherical cell is stable with respect to spherical disturbances.
Note that these results are similar to the 2D case \cite{AbuHamed2020}, where we find that the circular cell shape is stable concerning a small disturbance.

\section{Conclusion}
We perform the analysis of a minimal phase field model that is a 3D generalization of the 2D model
developed and investigated numerically in
\cite{Ziebert2012}, \cite{Ziebert2014}, and \cite{Ziebert2016} (a similar analysis of the latter model was done in
\cite{AbuHamed2020}). In this model the
order parameter $u$ is coupled with 3D polarization (orientation) vector field $\textbf{P}$ of the actin network. The
model is supposed to describe the 3D cell motility immersed in 3D ECM via actin based protrusion mechanism
\cite{Caswell+Zech2018}.

We considered the rotational symmetric case i.e., spherical shape interface, where we obtained a closed
ordinary differential equation describing the evolution of the radius \eqref{R(t)1}. We found the minimum value $\beta_c$ for the actin creation that is compatible with the existence of a stationary cell solution \eqref{b-c}. We found that when $\beta_c < \beta$, the circular cell can have some stationary radius, while in the case $\beta\leq\beta_c$ the cell shrinks until it disappears, which is meaningless in the context of cell dynamics. Also, we considered the general shape 3D cell dynamics. We found the
leading order solutions, \eqref{l1}-\eqref{l3}, and derived a closed integro-differential equation \eqref{gfd} governing the 3D cell
dynamics, which includes the normal velocity of the membrane, curvature, volume relaxation rate, and a parameter $\Omega$ determined by
the molecular effects of the subcell level. This result is similar to the 2D case \cite{AbuHamed2020}.

We found an equation of motion of the cell interface that can be written in the canonical form,

 \[
   v_n = -2D_u \mathcal{H} -  \tilde{V} + \Omega.
 \]

The stability analysis shows that the non-spherical shape and the motion of the cell as a whole cannot appear due to the development of a linear instability
of the spherical cell with the spherically symmetric polarization field localized near the cell boundary. In the framework of the considered model, the
transition to a non-spherical shape needs a finite-amplitude disturbance significantly changing the polarization field inside the cell. The analysis
of that transition, which can be carried out only numerically, is beyond the scope of the present paper.\\

\appendix\section{}\label{A}
\begin{figure}
  \centering
  \includegraphics[scale=0.3]{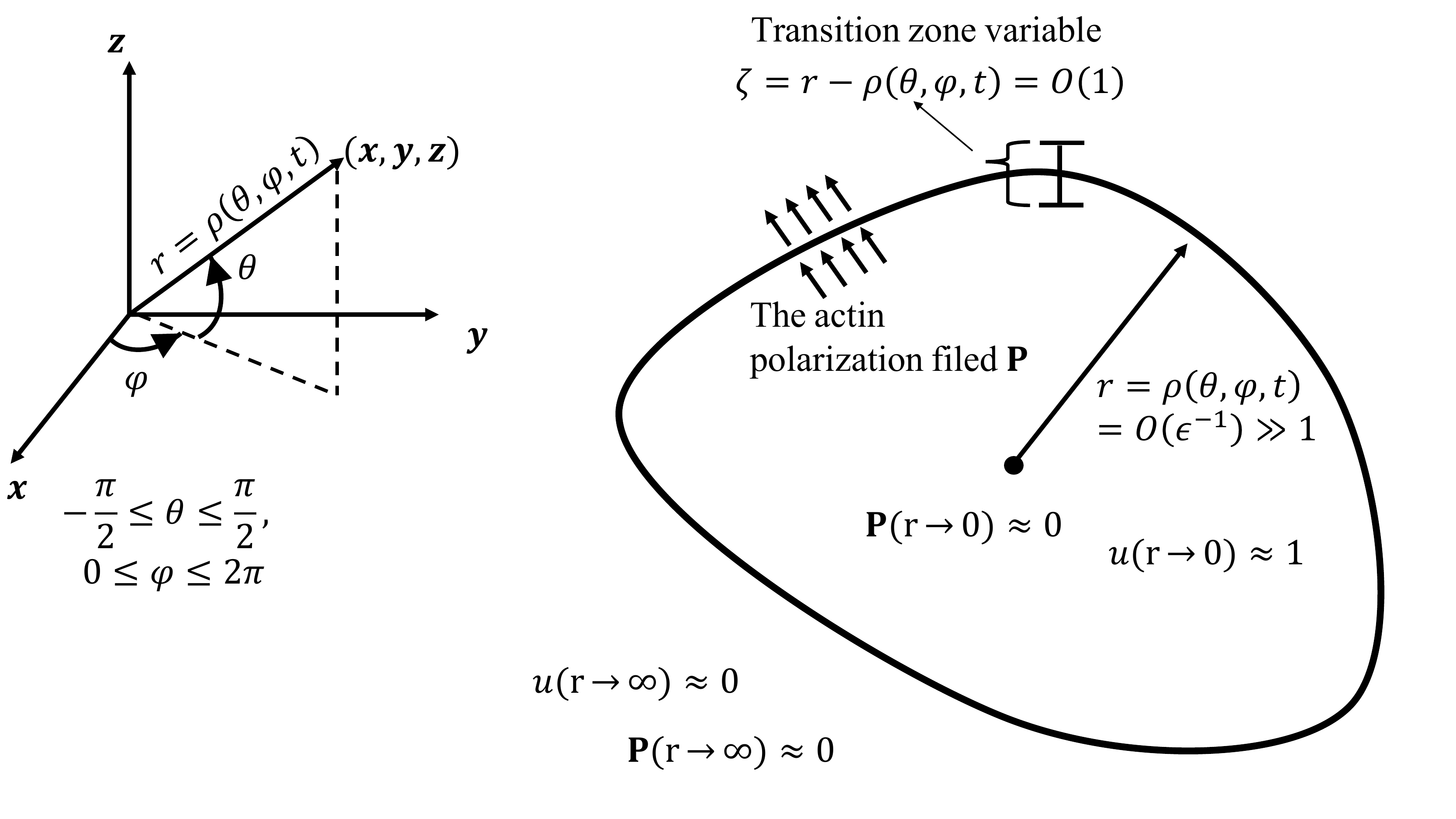}
  \caption{ A schematic description of the 3D cell membrane dynamics  in spherical coordinate system, and the boundary
conditions of the simplified model \eqref{Mod1}-\eqref{Mod6}.
    The thickness of the cell wall (i.e., the width of the transition zone, where $u(r,t)$ is changed from nearly 1 to nearly 0) is $O(1)$, and the cell
size is large. Therefore the ratio $\epsilon$ of the thickness of the cell wall to the size of the cell is small. } \label{schematic}
\end{figure}

Here we give an explicit expression for the mean curvature of a surface given in spherical coordinate description
$r=R(\ta,\ph,t)$, see Fig. \ref{schematic}. Following the definition \eqref{curv},\eqref{tensors},
one can calculate,
\begin{eqnarray}\label{}
  &&  \nabla \cdot \hat{n} = \Lambda \Bigg\{ \frac{2}{R} - \frac{R_{\ta\ta}}{R^2} + \frac{R_\ta \tan\ta}{R^2} - \frac{R_{\ph\ph}}{R^2 \cos^2\ta} \Bigg\}\nonumber\\
  && +\Lambda^3 \Bigg\{ \frac{1}{R^3}\left( R_\ta^2 + \frac{R_\ph^2}{\cos^2 \ta} \right)+ \frac{R_\ta^2 R_{\ta\ta}}{R^4} +\frac{2R_\ta R_\ph R_{\ta\ph}}{R^4 \cos^2 \ta}\nonumber\\
  &&+ \frac{R_\ph^2}{R^4 \cos^4 \ta } \left( \frac{1}{2} R_\ta  \sin2\ta +  R_{\ph\ph}  \right)    \Bigg\}
 \end{eqnarray}

\bibliography{FSD2.1.bbl}{}

\newpage

\end{document}